\documentstyle[11pt]{article}

\title{\bf On the initial condition for evolution of the
perturbative QCD Pomeron in the nucleus}
\author{M.A.Braun\\
Dep. of High Energy physics,
 University of S.Petersburg,\\
198504 S.Petersburg, Russia}
\def\beq{\begin{equation}}
\def\eeq{\end{equation}}
\def\noi{\noindent}
\begin{document}
\maketitle
\medskip
\noi{\bf Abstract.}
It is shown that subdominant terms found in the reggeized gluon diagram
technique, to be added to  Pomeron fan diagrams with the 3P interaction, can
be exactly taken into account by taking the initial condition for evolution
in the Glauber form. This demonstrates complete equivalence
in the high-energy limit of the
dipole picture and reggeized gluon technique not only on the leading level
but also on the subleading level.

\section{Introduction}
As shown in ~\cite{bra1,BLV} the equation for propagation
of the perturbative QCD pomeron in the nucleus
(BK equation, ~\cite{bal, kov}) corresponds to the summation
of  fan diagrams constructed of the pomeronic Green functions
connected by the splitting triple pomeron vertex $V$ first introduced
in the dipole picture by A.Mueller and B.Patel ~\cite{MP},
afterwards derived for arbitrary number of colours
from the  reggeized gluon diagrams by J.Bartels and M.Wuesthoff  ~\cite{BW}
and simplified for the mulicolour case by G.P.Vacca and the author ~\cite{BV}.
In the latter approach, however the contribution of the triple Pomeron
interaction does not exhaust the total amplitude. Apart from it,
there appears a contribution  from the so-called 'reduced' or 'reggeized'
term which is given by a single Pomeron exchange coupled to all
colour dipoles in
the multi-nucleon target. This contribution was explicitly introduced
for 4 gluons in ~\cite{BW} and for 6 gluons in ~\cite{BE}.
The reggeized contribution is obviously subdominant at high energies, so
taking it into account in the  BK equation is not really justified.
However already at an earlier stage of the discussion of the BK equation
we conjectured that one can  reproduce the reggeized contribution by
suitably generalizing the initial condition for the evolution. This idea
was actually initiated by our discussion with Yu.Kovchegov and E.Levin of
the Glauber-like initial condition proposed in ~\cite{kov}. In particular
E.Levin rightfully noted that multiple two-
gluon exchanges are not contained in the triple Pomeron interaction diagrams,
and should be added via glauberization of the single-nucleon initial
condition ~\cite{lev}. This problem is strictly speaking  not
important for the formulation of the BK equation nor for its actual solution,
since it refers to  subdominant contributions and, as calculation show,
does not practically influence the solution already after evolution over
some several of units of rapidity. However it is of some interest to see if
the correspondence between the dipole picture and reggeized gluon technique
can be extended also to this subdominant level. Also it may have some
practical effect when one is interested in the evolution to not so high
rapidities, when the result preserves certain trace of the chosen
initial condition.

In this paper we prove that  the reggeized terms which are found in the
reggeized gluon technique, to  be added to the triple pomeron
interaction terms, are indeed exactly reproduced by choosing the Glauber-like
initial condition in the BK equation.

\section{Initial condition for  evolution}
The amplitude for the scattering of some projectile on the nucleus
at rapitity $y$ and fixed impact parameter $b$ can be presented as
\beq
{\cal A}(y,b)=2is\int d^2r \rho(r)\Phi(r,y,b),
\eeq
where $\rho(r)$ is the colour dipole density in the projectile and
$\Phi(r,y,b)$ is the amplitude for the scattering of a dipole on the
nucleus. In the reggeized gluon diagram technique it is given by a sum of
all fan diagrams constructed from the BFKL Green functions $G$ and triple
Pomeron vertex $V$  for the spliiting of a Pomeron in two. For our purpose we
do not need the explicit form of both, which can be found in the cited
references. What we need is the lowest order term, corresponding to the
single Pomeron exchange:
\beq
\Phi_1(r,y,b)=\int d^2r'G(y,r,r')AT(b)\rho_N(r'),
\eeq
where $G(y,r,r')$ is the (forward) BFKL Green function, $AT(b)$ is the
transverse nuclear density at impact parameter $b$ and $\rho_N(r)$ is
the colour dipole density of the nucleon.

To compare with the results obtained by the reggeized gluon diagram technique
we introduce a Pomeron coupled to the projectile
\beq
P(y,r)=\int d^2r'\rho(r')G(y,r',r)
\eeq
and present the reduced amplitude ${\cal A}/(2is)$ as
\beq
\tilde{\cal A}(y,b)\equiv
\int d^2r \rho(r)\Phi(r,y,b)=\int d^2rP(y,r)AT(b)\rho_N(r).
\eeq
Next we amputate the Pomeron $P(r)$ from the target side, that is
separate the gluon propagators $1/k^4$ from its Fourier transform including
them into the target impact factor, to write  (4) as
\beq
\tilde{\cal A}(y,b)=\int d^2r D(y,r)\tau(r,b).
\eeq
Here $D(y,r)$ is the amputated Pomeron coupled to the projectile and
$\tau(r,b)$ is determined as a  Fourier transform:
\beq
\tau(r,b)=AT(b)\int\frac{d^2k}{(2\pi)^2k^4}e^{ikr}\rho_N(k).
\eeq

At first sight (6) is infrared singular since the integral is divergent
at $k\to 0$.
However this diveregence is spurious. In fact in (6) we can subtract
unity from the exponential, presenting $\tau(r)$ as
\beq
\tau(r,b)=\tau(0,b)+\tilde{\tau}(r,b),
\eeq
where
\beq
\tilde{\tau}(r)=\int\frac{d^2k}{(2\pi)^2k^4}\Big(e^{ikr}-1\Big)\rho_N(k)
\eeq
and is infrared finite and $\tau(0)$ is infrared divergent but
independent of $r$. Since
\beq
\int d^2rD(y,r)=0,
\eeq
in (5) we can drop the first term of (7) and write
\beq
\tilde{\cal A}(y,b)=g^2\int d^2r\tilde{D}(r,y)\tilde{\tau}(r,b),
\eeq
where we also extracted $g^2$ from $D$ by writing
\beq
D(r,y)=g^2\tilde{D}(r,y).
\eeq
Expression (10) is obviously infrared finite.

Our next aim will be to see if appropriately changing in (10) the
initial condition given by the impact factor $g^2\tilde{\tau}(r)$
we can reproduce the reggeized terms found in the analysis of the
reggeized gluon diagrams. Arguments $y$ and $b$ will always be fixed
in the following and we suppress them altogether.

\section{Two scattering centers}
We start from the contribution from the exchange of 4 reggeized gluons
$D_4$ which has been studied in detail in ~\cite{BW,bra2,BV}. It has been
found that the total contribution can be separated in two terms: a term with
the triple Pomeron interaction and a reggeized term $D^{(R)}_4$, which
corresponds to a single Pomeron exchange coupled to both traget centers.
Naturally at asymptotic energies the reggeized term is subdominant, since
if $\Delta$ is the BFKL intercept (in fact the Pomeron intercept minus unity)
$D^{(R)}_4$ behaves as $\exp (\Delta y)$,
whereas the triple Pomeron term behaves as $\exp (2\Delta y)$.

The contribution of the reggeized term to the amplitude is given by
\beq
\tilde{\cal A}^{(R)}_4=\frac{1}{2}\int d^2r_1d^2r_2D^{(R)}_4(r_1,r_2)\tau(r_1)
\tau(r_2),
\eeq
where $D^{(R)}(r_1,r_2)$ is the Fourier transform of the reggeized
part for two forward Pomerons:
\beq
D^{(R)}_4(r_1,r_2)=\int\frac{d^2k_1}{(2\pi)^2}\frac{d^2k_2}{(2\pi)^2}
e^{-ik_1r_1-ik_2r_2}D^{(R)}_4(k_1,-k_1,k_2,-k_2).
\eeq
Here $D^{(R)}_4(k_1,k_2,k_3,k_4)$ is the reggeized term as a function of
gluon momenta in the high-colour limit (see ~\cite{BW,BV}):
\beq
D_4^{(R)}(1,2,3,4)=\frac{1}{2}g^2\Big(\sum_{i=1}^4D(i)
-\sum_{i=2}^4D(1i)\Big),
\eeq
in the shorthand notations $k_1\equiv 1$, $k_1+k_2\equiv 12$ etc.
For our forward case, taking into account that $D(0)=0$, we find
\beq
D_4^{(R)}(k_1,-k_1,k_2,-k_2)=\frac{1}{2}g^2
\Big(2D(k_1)+2D(k_2)-D(k_1+k_2)-D(k_1-k_2)\Big).
\eeq
Performing the Fourier transform we find
\beq
D^{(R)}_4(r_1,r_2)=\frac{1}{2}g^2
\Big(2\delta^2(r_2)D(r_1)+2\delta^2(r_1)D(r_2)-g^2\delta^2(r_1+r_2)D(r_1)
-g^2\delta^2(r_1-r_2)D(r_1)\Big).
\eeq
Putting this into (12) we get
\beq
\tilde{\cal A}^{(R)}_4=\frac{1}{2}g^2\Big\{2\tau(0)\int d^2rD(r)\tau(r)-
\int d^2rD(r)\tau^2(r)\Big\},
\eeq
or, presenting $\tau(r)$ and $D(r)$ acording to (7) and (11) ,
\beq
\tilde{\cal A}^{(R)}_4=-\frac{1}{2}g^4\Big\{
\int d^2r\tilde{D}(r)\tilde{\tau}^2(r)+\tau^2(0)\int d^2r\tilde{D}(r)\Big\}.
\eeq
The last term is equal to zero, so that finally the contribution to
the amplitude from the reggeized term is
\beq
\tilde{\cal A}^{(R)}_4=
\int d^2r\tilde{D}(r)\Big[-\frac{1}{2}\Big(g^2\tilde{\tau}(r)\Big)^2\Big].
\eeq

Comparing with (10) we see that we shall reproduce this contribution
if in the the lowest order term we substitute
\beq
g^2\tilde{\tau}(r)\to g^2\tilde{\tau}(r)-\frac{1}{2}\Big(g^2\tilde{\tau}(r)
\Big)^2,
\eeq
which exactly corresponds to glauberizing the initial condition:
\beq
g^2\tilde{\tau}(r)\to 1 -e^{-g^2\tilde{\tau}(r)}.
\eeq

\section{Any number of scattering centers}
The preceding  derivation can be easily generalized to any (even) number
of gluons $n=2p$. Using the expression for the vertex connecting
a virtual photon to $n$ gluons arranged in colourless pairs
\{12\},\{34\},...\{(n-1)n\},
found in the linit $N_c\to\infty$ in
~\cite{braold}, one obtains in this limit
\beq
D^{(R)}_n(k_1,...k_n)=
-\frac{g^{n-2}}{2^p}\int d^2rD(r)\prod_{i=1}^{n}\Big(e^{ik_ir}-1\Big).
\eeq
One can check that this expression coincides with Eq. (14) for $n=4$ and
with $D^{(R)}_6$  found in ~\cite{BE} for general $N_c$ if one takes
$N_c\to\infty$.
In our case we have to take each target at zero momentum transfer, that is
$k_1+k_2=k_3+k_4=...=k_{n-1}+k_n=0$ with  $p$ momenta $k_1,k_3,...k_{n-1}$
as independent variables. To couple to the target we transform (22) to the
coordinate space $r_1,r_3,...r_{n-1}$:
\beq
D^{(R)}_n(r_1,r_3,...r_{n-1})=
-\frac{g^{n-2}}{2^p}\int d^2rD(r)\prod_{i=1}^{p}\Big[
\frac{d^2k_{2i-1}}{(2\pi)^2}\Big(2-e^{ik_{2i-1}r}-e^{-ik_{2i-1}r}\Big)\Big].
\eeq

Contributions which come from 2 in one or several bracket factors
lead to terms containing  $\delta^2(r_i)$  or several such factors.
After integration with the target factors $\tau(r_i)$ they will produce
infrared divergent terms containing $\tau(0)$. However we know
from the start that such terms
should cancel, since the initial expression (22) is obviously integrable
with the target factor in the infrared \footnote{We are indebted to G.P.Vacca
who drew our attention to this point}. Knowing this we
may drop 2 from the brackets in (23). Furthermore both terms remaining in the
brackets will obviously
give the same contribution after integration with the target factor, since
the change ${\bf r}\to -{\bf r}$ does not change the latter. So effectively
we get $2^p$ equal terms and the final result is easily found to be
\beq
D^{(R)}_n(r_1,r_3...r_{n-1})=
-(-1)^pg^{n-2}D(r_1)\prod_{i=2}^{p-1}\delta^2(r_{2i-1}-r_1)
\eeq
(at $n=2$ there are no $\delta$-functions and $D^{(R)}_2(r)=D(r)$
as expected).
Integrating this expression with $p$ target factors  and taking into account
the symmetry factor $1/p!$ we obtain
\beq
\tilde{\cal A}^{(R)}_n=-(-1)^p\frac{1}{p!}g^{2p-2}\int d^2r D(r)\tau^p(r).
\eeq
Eliminating the infrared divergent contributions from
powers of $\tau(0)$, which, as
mentioned, should be completely cancelled out, and passing to $\tilde{D}$ we
finally find
\beq
\tilde{\cal A}^{(R)}_{2p}=-(-1)^p\frac{1}{p!}g^{2p}\int d^2r
\tilde{D}(r)\tilde{\tau}^p(r).
\eeq
Summing the contributions for all $p$ we obtain the Glauber expression under
the sign of integral over $r$
\beq
\tilde{\cal A}^{(R)}
=\sum_{p=1}\tilde{\cal A}^{(R)}_{2p}
=\int d^2r \tilde{D}(r)\Big(1-e^{-g^2\tilde{\tau}(r)}\Big),
\eeq
which means that indeed all the reggeized contribution can be adequately
taken into account by glauberizing the initial condition according to (20).

\section{Conclusions}
We have shown that subdominant terms found in the reggeized gluon diagram
technique, to be added to the Pomeron fan diagrams with 3P interaction, can
be exactly taken into account by taking the initial condition for evolution
in the Glauber form. This demonstrates  complete equivalence of the
dipole picture and reggeized gluon technique not only on the leading level
in the high-energy limit but also on the subleading level.

As mentioned, glauberization of the initial condition does not make any
difference after evolution by several units of rapidity: the BK equation
rapidly forgets the details of the initial condition. However the difference
is certainly felt at lower stages of evolution, which may be of
importance in practical applications.

\section{Acknowledgements}
The author has benefited by numerous discussions with J.Bartels, Yu.Kovchegov,
E.Levin and G.P.Vacca. This work was supported by the NATO grant
PST.CLG.980287.

\end{document}